\documentclass[pra,amsmath,twocolumn,showpacs]{revtex4-1}
\usepackage{csquotes}
 
\usepackage{blindtext}
\usepackage{graphicx}
\usepackage{amsmath,graphicx}
\usepackage{xcolor}
\usepackage{commath}
\usepackage{float}
\usepackage{fullpage}
\usepackage{csquotes}
\usepackage{amsmath,amssymb,stmaryrd}
\usepackage{dsfont}
\usepackage{braket}
\def\la{{\langle}}
\def\e{\enquote}
\def\u{\hat U}
\def\A{\mathcal A}
\def\AA{{\bf   A }}
\def\AAA{{\bf A}}
\def\PP{\tilde P^S}
\def\Z{Z}
\def\PPP{\tilde P}
\def\B{\hat B}

\def\F{\hat F}

\def\up{\uparrow}

\def\h{\hat H}

\def\q{\quad}

\def\om{\omega}

\def\g{\color{green}}

\def\n{\\ \nonumber}

\def\ra{{\rangle}}

\def\h{\hat{H}}

\def\g{{\gets}}

\def\e{\enquote}

\def\I{{\text {Im}}}
\def\R{\text{Re}}
\def\dx{\Delta x}
\def\up{\uparrow}
\def\dn{\downarrow}
\def\q{\quad}
\def\n{\\ \nonumber}

\begin{document}
\title{Inaccurate (weak) measurements classical and quantum}
\author{D. Sokolovski$^{a,b,c}$}
\email{dgsokol15@gmail.com}
\author{D. Alonso$^{d}$}
\author{S. Brouard$^{d}$}
\affiliation{$^{a}$ Departmento de Qu\'imica-F\'isica, Universidad del Pa\' is Vasco, UPV/EHU, 48940, Leioa, Spain}
\affiliation{$^{b}$ IKERBASQUE, Basque Foundation for Science, Plaza Euskadi 5, 48009, Bilbao, Spain}
\affiliation{$^{c}$ EHU Quantum Center, Universidad del Pa\' is Vasco, UPV/EHU, 48940, Leioa, Spain}
\affiliation{$^{d}$ Departamento de F\'isica y IUdEA, Universidad de La Laguna,La Laguna,Tenerife, Spain}
\begin{abstract}
\begin{center}
{\bf ABSTRACT}
\end{center}
\noindent
We consider highly inaccurate measurements made on classical stochastic and quantum systems. In the quantum case such a \e{weak} measurement preserves coherence between the system's alternatives. We demonstrate that in both cases the information about the scenario realised in each individual trial is lost. However, ensemble parameters such as classical path probabilities, and quantum quasi-probabilities can be extracted from the obtained statistics. In both cases causality ensures that additional post-selection only redistributes individual outcomes between the system's final states. Quantum quasi-probabilities may change sign, which allows for anomalously large meter's (pointer's) reading for some final states. These, we show, result from mere \e{reshaping} of a broad distribution obtained earlier, and provide no \e{experimental evidence} of quantum variables taking, on rare occasions, exceptionally large values. 
\end{abstract}
\today
\maketitle
\section{Introduction}
The problem of establishing intermediate condition 
of a quantum system making a transition between 
given initial and final states has been a source of contradiction 
for several decades (see, e.g., \cite{SPIN100}). 
In order to alter the transition probability as little as possible 
one can try to use a highly inaccurate pointer, such that 
the uncertainty in its initial position greatly exceeds the 
displacements it may undergo in the course of the measurement. 
\newline
The difficulty with such an approach is well known in the text book 
quantum theory of \cite{FeynL}. The system can be seen as reaching 
its final state vie virtual (Feynman) paths alongside which the 
quantity takes definite values. An inaccurate pointer is designed to leave
interference between them intact. The Uncertainty Principle \cite{FeynL}
forbids knowing the path chosen by the system and, by extension, 
the value of the variable has taken. Among the well known quotes on the subject are Bohm's mention of \e{absurd results}
which can occur, should interference not be destroyed \cite{Bohm}, and Feynman's 
\e{blind alley} warning agains rationalising, in classical-like terms, the manner in which 
a quantum system travels interfering routes \cite{FeynC}.
\newline
It must be noted that  a different approach has been proposed in \cite{AH90}.
 In the conventional narrative, quantum mechanics calculates the
transition amplitudes as matrix elements of unitary operators between the states 
of interest, so one does not have to worry about the collapse of the wave function 
at the end of the experiment. The authors of \cite{AH90} literally doubled down on the
role of the wave function, by adding to the description the backwards-evolving final state.
The method was followed on under the name time-symmetric quantum mechanics (TSQM) (see, e.g.,
\cite{Ah2T}). 
Interpretational value  (no new results are predicted by the TSQM  \cite{AH90}) of this picture was questioned in \cite{WVcrit1}.
More recent critique was given in \cite{WVcrit2}, where the reader will also find further relevant references from the extensive literature existing on the subject. 
We share the concerns of  \cite{WVcrit1}, \cite{WVcrit2}, and will henceforth remain within 
the conventional framework of \cite{FeynL} outlined in more detail in  \cite{DSS}.
\newline
 Having thus stated our preference, we proceed to the purpose of this paper.
 Consider a pointer whose initial position is uncertain.
 The pointer is employed to measure a variable $\B$ of 
 a quantum system, previously prepared (pre-selected) and later found (post-selected) 
 in known initial and final states.  If the initial (Gaussian)  distribution of the pointer's
 position is  sufficiently broad, the final distribution of its (readings) 
is  given by the same Gaussian, displaced as a whole by the real part of a complex 
 quantity known as \e{the weak value (WV) of $\B$} \cite{SPIN100}, \cite{WVrev}. 
 It is easy to show that one can always find a suitable pair of the system's
 initial and final states so as to give the WV of $\B$  and, therefore, 
 to the pointer's shift any desired value \cite{DSS}. 
 Earlier we related the controversies surrounding the weak values
 to the predictive power erroneously attributed to quantum probability amplitudes, 
 and ultimately to the Uncertainty Principle (see. e.g.,  \cite{DSpla1}).
Here we consider the problem from a different perspective, and undertake to answer two important questions.
 \newline
* Can the shift of an inaccurate pointer be interpreted as a unique well defined value 
of $\B$, occurring each time the system makes the transition? The question is not new. 
A historic perspective on the single-system vs. ensemble property controversy
can be found in \cite{WVcrit2}.
\newline
* For an unlikely transition the shift can be very large. Does this mean
that, on rare occasions, quantum mechanics allows a variable  to
take anomalously large, or otherwise unusual, values?
The authors of \cite{SPIN100} exercise caution and talk about result of measuring 
 spin-1/2 \e{turning out to be $100$}. Reference \cite{Ah2T} speaks about the 
 discovery of \e{new quantum effects} related to this large value. 
Plainly, more clarity is desirable, and we will seek it in what follow. 
\newline
In doing so it will be useful to consider first a classical statistical model, 
in order to focus on the essentially quantum aspects of the problem later. 
The lack of clarity about the origin and status of quantum WV 
has lead in the past the authors of  \cite{Ferry} to an erroneous \cite{Comm1} claim
that anomalous WVs may exist in a purely classical environment. 
We will revisit this particular controversy in due course. 
\newline
The rest of the paper is organised as follows. 
In Section II we will review a useful mathematical property.
Sections III and IV discuss highly inaccurate classical 
measurements with and without post-selection, 
one's ability to reconstruct the path probabilities, 
and the classical causality principle.
Sections V and VI contain a similar analysis of inaccurate
(weak) quantum measurements. In Sect. VII
we ascribe quasi-probabilities to the system's virtual 
paths. Sect.VII contains a simple example. 
Sect. IX briefly discusses the meaning and use 
of negative probabilities in quantum theory. 
In Sect. X we discuss the reshaping mechanism 
producing the anomalous readings in the quantum case,
Sect. XI  gives another simple example.
Our conclusions and discussion are in Sect. XII.  
\section{An asymptotic property}
We begin with a curious mathematical property. 
For a smooth real function $G(s)$, $-\infty <\partial_x G(x)< \infty$ the following
is true in the limit where  the scaling parameter $\dx$
tends to infinity,  
\begin{eqnarray} \label{1}
\sum_{i=1}^N A_iG\left (\frac{x-B_i}{\dx}\right){\xrightarrow[ \dx \to \infty ] {}}
\left [\sum_{i=1}^N A_i\right ] G\left (\frac{x-z}{\dx}\right),\q\q \n
\end{eqnarray}
\begin{eqnarray} \label{2}
z\equiv 
\frac {\sum_{i=1}^N A_iB_i}{\sum_{i=1}^NA_i},\q\q\q\q\q\q\q\q\q\q
\end{eqnarray}
where $A_i$, and $B_i$, $i=1,...N$ are real, possibly negative numbers (see Appensix A). 
The convergence is point wise \cite{UDS}, as illustrated in Fig.1.
\begin{figure}[h]
\includegraphics[angle=0,width=6.5cm, height= 5.0cm]{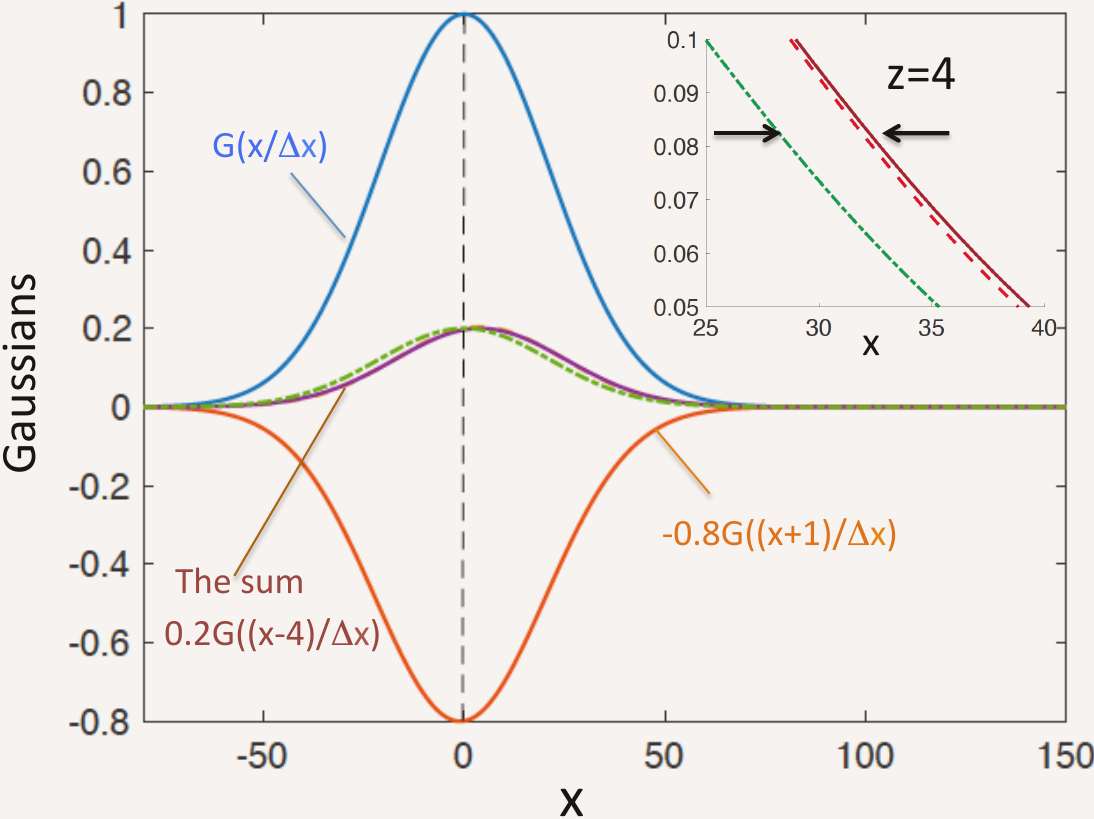}
\caption {Two broad Gaussians (\ref{5}) in the l.h.s. of Eq.(\ref{1}), $\dx=30$, $A_1=1$, $B_1=0$, and $A_2=-0.8$, $B_2=-1$, 
add up to a smaller Gaussian shifted to the right by approximately 4 units. The dot-dashed line shows the same Gaussian centred at $x=0$.
In the inset the added dashed line shows, for comparison, the Gaussian in the r.h.s. of Eq.(\ref{1}).
} 
\end{figure}
\newline
It follows further that for complex valued $A_i$ the absolute square of the sum 
in (\ref{1}) tends to a similar limit
  (see Appendix A)  
\begin{eqnarray} \label{3}
\left | \sum_{i=1}^N A_iG\left (\frac{x-B_i}{\dx} \right) \right |^2{\xrightarrow[ \dx \to \infty ] {}}
\n
\left | \sum_{i=1}^N A_i\right |^2  G^2\left (\frac{x-Z}{\dx}\right), 
\end{eqnarray}
where
\begin{eqnarray} \label{4}
Z\equiv \R \left [\frac{\sum_{i=1}^N A_iB_i}{\sum_{i=1}^NA_i}\right ].\q\q\q\q\q\q\q\q\q\q\q
\end{eqnarray}
To our knowledge, these limits were first reported in \cite{SPIN100} for the Gaussians,  
\begin{eqnarray} \label{5}
G(x)=\exp\left (-\frac{x^2}{\dx^2}\right). 
\end{eqnarray}
An early critique of \cite{SPIN100} suggested that whenever two Gaussians are added,
it should always be possible to recognise in the result the presence of the two constituent peaks \cite{Per}.
This, however, does not seem to be the case (cf. Fig.1, for relation to the Catastrophe 
Theory \cite{Cat} see \cite{UDS}).
Next we analyse the importance, if any, of Eqs.(\ref{1})-(\ref{4}) for highly inaccurate measurements made 
on both classical and quantum systems.
\section{Two-point classical measurements}
We start with the classical case.
Consider  a simple setup where a classical particle  
(one can think, if it helps,  of  a little ball rolling down a system of tubes or slides), 
introduced at $I$ into the system of routes shown in Fig.2a, 
is directed to the $i$-th  node, $i=1,2,..N$ with probability $P(i\gets I)$. From there it is sent to a  
receptacle (final state) $j$ $,j=1,2,..N$ with a probability $P(j\gets i)$. Thus, $N$ paths connect the initial state $I$ 
with a final state $j$ (see Fig.2b). A path $j\gets i \gets I$ is travelled with probability 
 \begin{eqnarray} \label{1a}
P(j\g i\g I)=P(j\gets i)P(i\gets I),   
\end{eqnarray}
so the net probability to reach the state $j$ is 
 \begin{eqnarray} \label{1ab}
P(j\g  I)=\sum_{i=1}^NP(j\g i\g I).
\end{eqnarray}
\begin{figure}[h]
\includegraphics[angle=0,width=7.5cm, height= 6.0cm]{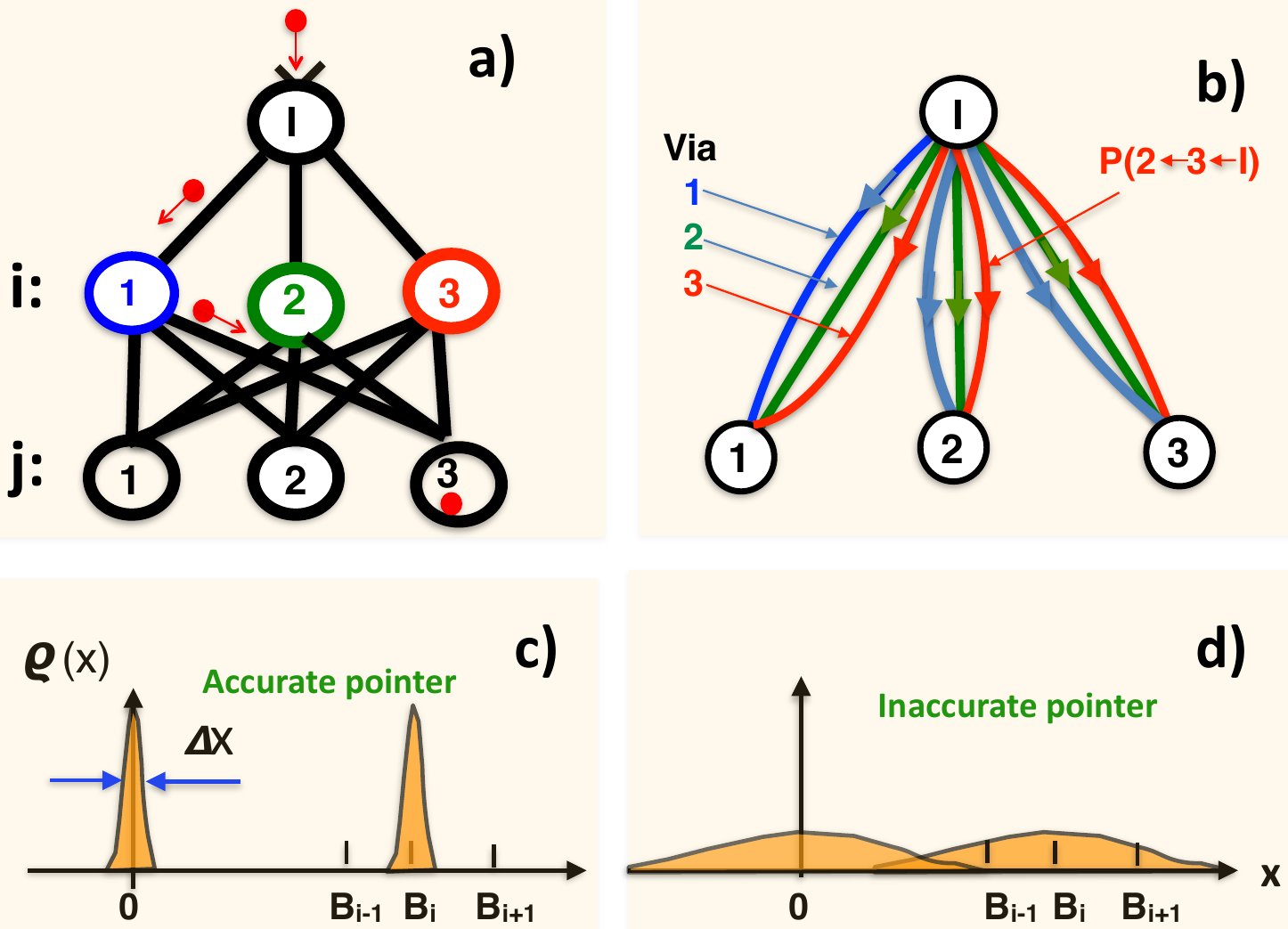}
\caption {a) Alice's classical setup ($N=3$).
The particle  passes through a state  $|i \ra$ on the way to its final state $j$.
b) Each of the nine paths
 available to the particle
is equipped with the the probability in Eq.(\ref{1a}).
c) initial uncertainty of an accurate pointer's position, $\dx$, 
is small compared to the differences $max |B_i|$.
d) for an inaccurate pointer one has $\dx >max |B_i|$.
} 
\end{figure}
\newline 
Alice-the-experimenter (\AAA) wants to measure a variable (a \e{property} \cite{Matz}) $B$ which takes the value $B_i$ if the particle
crosses the  $i$-th  node. For this purpose she uses a pointer with position $x_1$) 
which moves by $B_i$ units whenever $i$-th intermediate state is crossed. 
The pointer's initial position, not known exactly, is distributed around zero
according to 
 \begin{eqnarray} \label{3o}
\rho^I(x_1)=\pi^{-1/2}\dx_1^{-1}\exp(-x_1^2/\dx_1^2). 
\end{eqnarray}
The distribution of its final positions (readings), which can be determined accurately,  is, therefore, given by
 \begin{eqnarray} \label{3a}
\rho(x_1)=   \sum_{i=1}^N P(i\g  I)\rho^I(x_1-B_i).
\end{eqnarray}
If the pointer is {accurate}, $\dx_1 \to 0$, $\rho(x_1)=   \sum_{j=1}^N P(i\g  I)\delta(x_1-B_i)$, 
where $\delta(x)$ is  the Dirac delta (see Fig.2c).  Thus, with all $B_i$ distinct,  \AA always knows the ball's position. 
\newline
We are, however, mostly interested in the case where \AA has at her disposal only a {highly  inaccurate} pointer 
 with 
 $\dx_1 >>\text{max} |B_i|$ (we will write simply $\dx_1 \to \infty$). If so, by Eq.(\ref{1}), the distribution (\ref{3a}) is a single broad Gaussian
 \begin{eqnarray} \label{4a}
\rho(x_1){\xrightarrow[ \dx _1\to \infty ] {}} 
\rho^I\left (x_1-y(B)\right),\n   
y(B)\equiv \sum_{i=1}^N B_i P(i\g  I).
\end{eqnarray}
With all the Gaussians $\rho^I(x-B_i)$ in (\ref{3a}) overlapping, \AA cannot know the particles's location. However, after many trials
she can evaluate the average value of the variable $B$, $y(B)$. If $B$ probes the particle's  presence 
in the   $i_0$-th node , $B_i= \delta_{ii_0}\equiv \pi_{i_0}$, \AA estimates in this manner the probability $P(i_0\g I)$.
\newline
Finally, Eq.(\ref{4a}) may look as if a measurement has been made on a system where $B$ has a unique well defined 
value $y(B)$. This is, of course,  not the case, as $y(B)$ is an average over many trials. 
\section{Inaccurate classical measurements with \e{post-selection}}
Alice can extend her experiment, and use a pointer with position 
$x_2$ to measure a variable $F$, which takes a  value $F_j$ 
if the particle ends up in its $j$-th final state. 
The final distribution of the  pointers' reading  is given by 
 \begin{eqnarray} \label{2b}
\rho(x_1,x_2)= \sum_{i,j=1}^NP(j\g i\g I)\times \n
\rho^I_1(x_1-B_i)\rho^I_2(x_2-F_j),\q\q
\end{eqnarray}
where $\rho_k^I(x_k)=\pi^{-1/2}\dx_k^{-1}\exp(-x_k^2/\dx_k^2)$, $k=1,2$.
Again, with both pointers accurate, $\dx_1,\dx_2\to 0$, 
$\rho^I_{1,2}(x_{1,2})$ tend to $\delta(x_{1,2})$ and
any pair of  readings $(x_1\approx B_i,x_2\approx F_j)$
allow \AA to identify the path $j\g i\g I$ travelled in the trial. 
\newline
However, we are interested in the case $\dx_1\to \infty$, $\dx_2\to 0$, 
whereby, in a given trial, the detailed route remains unknown.
By (\ref{1}) we have
  \begin{eqnarray} \label{3b}
\rho(x_1,x_2){\xrightarrow[ \dx_1 \to \infty, \dx_2 \to 0 ] {}}\q\q\q\q\q\q\q\q\n
\sum_{j=1}^N P(j\gets I) \rho^I_1(x_1- z_j(B))\delta(x_2-F_j), \q\q \n
z_j(B)
\equiv \frac{\sum_i B_i P(j\g i\g I)}{\sum_i P(j\g i\g I)}.\q\q\q\q\q\q\q
\end{eqnarray}
Alice may choose to select only the cases where $x_2\approx F_{j}$ and the ball 
is \e{post-selected} in the final state $j$. 
If so, the shift (position of the peak, or of the centroid) of the broad Gaussian 
in (\ref{3b}) coincides with the mean value of $B$, $z_j(B)$, calculated  with the 
probabilities (\ref{1a}) of the paths ending in $j$-th final state. 
For the special choice $B_i=\pi_{i_0}$, 
  \begin{eqnarray} \label{4b}
z_j ( \pi_{i_0})= \frac{ P(j\g i_0\g I)}{ P(j\g  I)}
\end{eqnarray}
yields the  probability of passing through $i_0$ on the way to the final state $j$.
\newline
We note (this will be important for future discussion of the quantum case) that 
Alice, who can determine both $z_j ( \pi_i) $ [by measuring the shift of the Gaussian in (\ref{3b})] and 
$P(j\g  I)$ [by counting the arrivals in $j$] is able to recover the probabilities $P(j\g i\g I)$ for all paths 
in Fig.2b, 
  \begin{eqnarray} \label{5b}
P(j\g i\g I)=z_j ( \pi_{i})P(j\g  I).
\end{eqnarray}
\newline
Finally, since $\sum_j P(j\g I)=1$, integration of Eq.(\ref{2b}) over $dx_2$ recovers the distribution (\ref{3a})
 obtained in the experiment in which 
the future state of the system is not known, or not controlled,
 \begin{eqnarray} \label{6b}
\int \rho(x_1,x_2)dx_2= \sum_{i=1}^NP(i\g I) \rho_I^I\left (x_1- B_i\right ). 
\end{eqnarray}
Equation (\ref{6b}) illustrates  the causality 
principle, built into classical statistics. 
Summation over the final readings {\it must}
recover the distribution (\ref{3a}) obtained 
in the shorter measurement of the previous Section. 
Were this not the case, \AAA's future decision to post-select
could be known in advance, and this should not be possible. 
\section{Two-point quantum measurements}
While classical statistic requires no introduction, the purpose of quantum theory  needs to be stated with sufficient clarity.
We take the main points from \cite{FeynL} (for more detail see also \cite{DSS}). 

(i) For each possible scenario the theory provides a complex valued probability amplitude, whose absolute 
square gives the probability of the scenario to occur.

(ii) If two scenarios leading to the same outcome cannot be distinguished, their amplitudes must be added. 
Otherwise one must add the corresponding probabilities.

(iii) Amplitudes of scenarios leading to different final outcomes are never added. 
\newline
Consider a quantum system, whose distinguishable conditions are in correspondence with orthogonal states  in an $N$-dimensional Hilbert
space. For such a system a possible scenario  is to start in a state $|I\ra$ at $t=0$, be in one of the states 
$|b_i\ra$, $\la b_i|b_{i'}\ra = \delta_{i'i}$, $i=1...N$, at $t_1>0$, and end up in one of the orthonormal states $|f_j\ra$
$j=1,...,N$ at $t_2>t_1$. 
\begin{figure}[h]
\includegraphics[angle=0,width=7.5cm, height= 3.5cm]{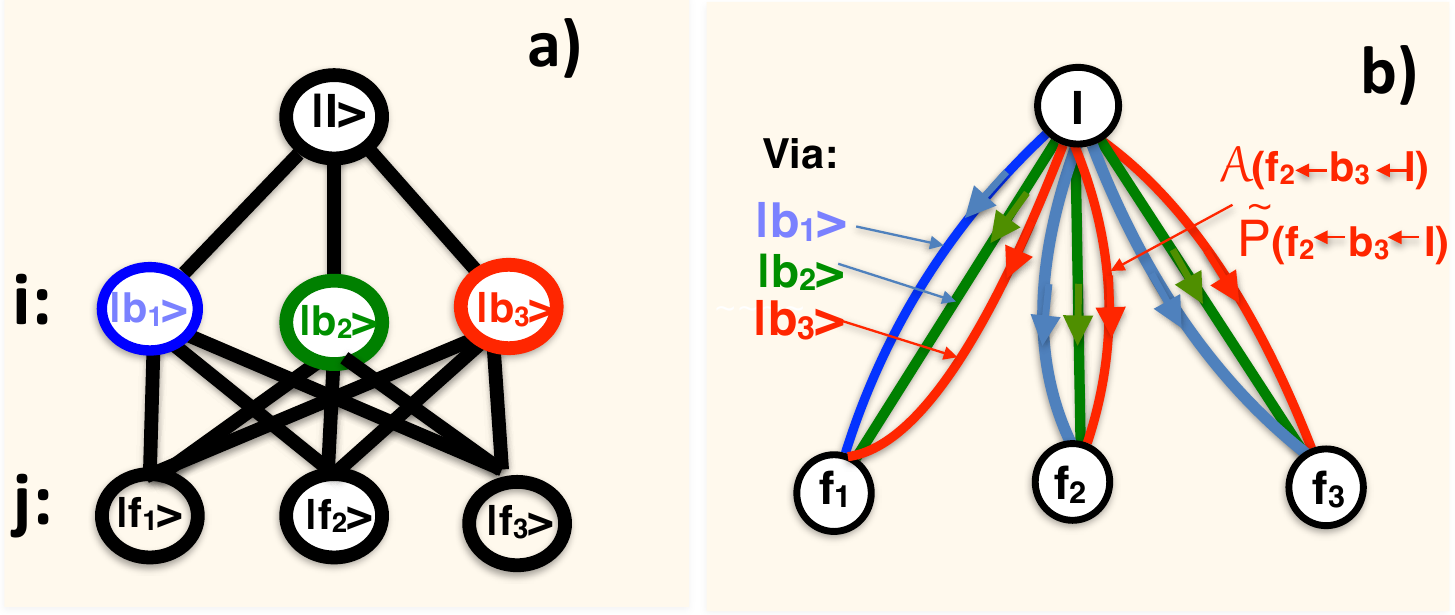}
\caption {a) Alice's quantum setup  ($N=3$).
A system passes through a state  $|b_i \ra$ on the way to its final state $|f_j\ra$.
b) Each of the nine virtual paths available to the system
is equipped with with the amplitude in Eq.(\ref{1c}). 
Quasi-probabilities (\ref{1e}) can also be ascribed without destroying 
coherence between the paths (see Sect.VI)}
\end{figure}
For each scenario the theory provides a probability amplitude (we use $\hbar=1$)
  \begin{eqnarray} \label{1c}
\A^{S}(f_j\g b_i\g I)=\A^{S}(f_j\g b_i)\A^{S}(b_i\g I)
\equiv\n
\la f_j|\u^S(t_2-t_1)|b_i\ra\la b_i|\u^{S}(t_1)|I\ra,\q \q  
\end{eqnarray}
where 
$\u^{S}(t)\equiv exp(-i\h^{S}t),$
and $\h^{S}$ is the system's  Hamiltonian,  chosen, for simplicity, to be time-independent.
\newline 
The simplest question \AA may ask is \e{What is the probability of a system prepared in $|I\ra$ to be found  in $|b_i\ra$ at $t=t_1$?} 
By (ii), there are $N$ scenarios (paths) $b_i \g I$, which lead to distinguishable final outcomes. 
By (iii), the paths can be endowed with probabilities, 
  \begin{eqnarray} \label{3c}
P^S(b_i\g I)=|\A^S(b_i\g I)|^2, \q j=1,2...N. 
\end{eqnarray}
As in the classical case, the existence of the transition probabilities (\ref{3c}), which  follows from the rules (i)-(iii), does not seem to depend on the presence of a measuring device.
\newline
However, Alice, unable to observe the system directly, may need  to rely on an instrument, accessible to her senses. 
Thus, she employs a von Neumann pointer (P)\cite{vN} whose position,  $x_1$, can be determined accurately just after it couples to the
system at $t=t_1$.
(Technically speaking, she also needs another pointer to record the system's initial state $|I\ra$, but there is no need to mention it explicitly.)
 The pointer, designed to measure an operator 
   \begin{eqnarray} \label{4ca}
 \B=\sum_{i=1}^N|b_i\ra B_i\la b_i|,
 \end{eqnarray}
  is  prepared in a Gaussian state $|G_1\ra$, 
  \begin{eqnarray} \label{4c}
\la x_1|G_1\ra ={\pi}^{-1/4} \Delta x_1^{-1/2}\exp\left (-\frac{x_1^2}{2\dx_1^2}\right)\equiv G_1(x_1), \q\q 
\end{eqnarray}
so its initial distribution,
  \begin{eqnarray} \label{4ca}
\rho_1^I(x_1) = |G_1(x_2)|^2 ,
\end{eqnarray}
is  a Gaussian of width $\dx_1$.
Now \AA has to apply the rules (i)-(iii) to  the composite system+pointer. The composite can  
 make transitions from $|I\ra|G_1\ra$ to one of the distinguishable (orthogonal) states $|b_i\ra|x_1\ra$, 
%
Such is the property of the von Neumann pointer (see Appendix B) that
 the corresponding amplitude 
is given by
  \begin{eqnarray} \label{5c}
\A^{S+P}(b_i,x_1\g I,G)= \A^{S}(b_i\g I)G(x_1-B_i).\q\q
\end{eqnarray}
Again, by (iii), one can assign probabilities $P_i^{S+P}(x_1)=|\A^{S+P}(b_i,x_1\g I,G_1)|^2$.
To find the observed distribution of the pointer's readings, \AA needs to sum $P_i^{S+P}(x_1)$
over unobserved system's states $|b_i\ra$.  This yields expressions remarkably similar to its classical analogue,
(\ref{3a}). 
  \begin{eqnarray} \label{6c}
\rho(x_1,t_1)=  \sum_{i=1}^N  P^S(b_i\g I)\rho^I(x_1-B_i),
\end{eqnarray}
and, by (\ref{1}), 
 \begin{eqnarray} \label{7c}
\rho(x_1,t_1){\xrightarrow[ \dx _1\to \infty ] {}} 
\rho^I\left (x_1-Y(B)\right),
  \end{eqnarray}
  where
\begin{eqnarray} \label{8c}  
Y(B)\equiv \sum_{i=1}^N B_i P^S(b_i\g  I)=\la I(t_1)|\B|I(t_1)\ra,
\end{eqnarray}
and $|I(t_1)\ra\equiv \u^S(t_1)] |I\ra$, so $Y(b)$ is the usual 
average of the operator $\B$ in the evolved initial state $|I(t)\ra$,
The rest of the discussion can follow that of Sect.III.
One notes the consistency of the Feynman's rules (i)-(iii). Whenever
the theory is able to  predict probabilities for the {system in isolation} [cf. Eq.(\ref{3c})], 
the same probabilities appear when the rules are applied to the composite
{ system + measuring device} [cf. Eq.(\ref{6c})].
\section{Quantum measurements with {post-selection}}
The principal difference between the  classical and quantum cases 
becomes evident  when one considers scenarios where a quantum system arrives 
in a final state $|f_j\ra$ via  an intermediate state 
$|b_i\ra$. Whereas classical probabilities (\ref{1a}) are always available,
quantum theory predicates their existence 
on one's ability to \e{distinguish} between the alternatives. 
\newline
In practical terms, 
Alice, interested in the 
system making a three-point transition $|f_j\ra\g |b_i\ra \g |I\ra$, needs to employ 
two von Neumann pointers ($P_1$ and $P_2$),  designed to measure the system's operators
  \begin{eqnarray} \label{0d}
 \B=\sum_{i=1}^N|b_i\ra B_i\la b_i|\q
\text {and} \q\F=\sum_{j=1}^N|f_i\ra F_i\la f_i|, \q 
\end{eqnarray}
at $t=t_1$ and $t=t_2$, respectively.
The pointers are  
prepared in the Gaussian states [cf. Eq. (\ref{4c})].
\newline
Now the amplitude of the composite $S+P_1+P_2$ making a two-point transition 
$|f_j\ra|x_1\ra|x_2\ra \g |I\ra |G_1\ra |G_2\ra$ 
is given by (see Appendix B)
  \begin{eqnarray} \label{1d}
\A^{S+P_1+P_2}(f_j,x_1,x_2\g I,G_1,G_2)=\q \q\n
 \sum_{i=1}^N \A^{S}(f_j\g b_i\g I)
G_1(x_1-B_i)G_2(x_2-F_j).\q
\end{eqnarray}
By the rule (iii) the transition probability exists in the form 
  \begin{eqnarray} \label{2d}
P^{S+P_1+P_2}(f_j,x_1,x_2)
 =\n
 |\A^{S+P_1+P_2}(f_j,x_1,x_2\g I,G_1,G_2)|^2.
\end{eqnarray}
It is easy  to check that with two accurate
pointers, $\dx_1,\dx_2\to 0$, Eq.(\ref{2d}) reduces to 
$\sum_{i,j=1}^N|\A^{S}(f_j\g b_i\g I)|^2\delta(x_1-B_i) \delta(x_2-F_j)$.
With all $B_i$ and $F_j$ distinct, inspection of a pair of readings $x_1\approx B_i$, $x_2\approx F_j$, 
allows  \AA to identify the system's path $|f_j\ra \g |b_i\ra \g |I\ra$.  By (ii), the path
may be endowed with probability, $P^S(f_j\g b_i\g I)=|\A^{S}(f_j\g b_i\g I)|^2$.
The rest of the discussion can follow that of Sect. IV, and we leave at that. 
\newline
Again, we are interested  in
the case where,
 as in Section IV,  Alice employs one accurate,
$\dx_2\to 0$, and 
one highly inaccurate, $\dx_1\to \infty$,  pointer. 
In Eq.(\ref{1d}) the Gaussians with different shifts $B_i$ are almost identical, 
and no reading $x_1$ will help \AA to identify the path chosen by the system.
The rules (i)- (iii) give no recipe for assigning probabilities to the paths, so what, if anything, 
can be learnt about the system's condition at $t=t_1$?
\newline
Summation of (\ref{2d}) over $j$ yields the quantum analogue
 of the classical Eq.(\ref{2b})
 [we used Eqs.(\ref{1d}) and (\ref{3})]
  \begin{eqnarray} \label{3d}
\rho (x_1,x_2){\xrightarrow[ \dx_1 \to \infty,   \dx_2 \to 0] {}}\q\q\q\q\q
\n 
\sum_{j=1}^NP^{S}(f_j\gets I) \rho^I_1(x_1-\Z_j(\B)) \delta(x_2-F_j)
\end{eqnarray}
where 
 \begin{eqnarray} \label{3da}
P^{S}(f_j\gets I)=
 |\la f_j|\u^S(t_2)|I\ra|^2, 
\end{eqnarray}
and 
\begin{eqnarray} \label{4d}
Z_j(\B) \equiv \R\left [ \frac{\sum_{i=1}^N B_i \A^{S}(f_j\g b_i\g I)}{\sum_{i=1}^N\A^{S}(f_j\g b_i\g I)} \right ].
\end{eqnarray}
As in the classical case, the distribution of the first pointer's readings is a single Gaussian, displaced by some
sort of average, obtained not with the path probabilities [which, by (ii), need not exist], 
but with the path amplitudes [which are always available]. 
\newline
This result was first obtained in \cite{SPIN100}, in a different form. For $\h^S=0$, or $0< t_1<t_2$, 
the complex quantity in the square brackets of (\ref{4d}) can be written as 
\begin{eqnarray} \label{7d}
\frac{\sum_i \la f_j|b_i\ra B_i\la  b_i|I\ra  }{\sum_i \la f_j|b_i\ra \la  b_i|I\ra } = \frac {\la f_j| \B |I\ra} {\la f_j|I\ra}, 
\end{eqnarray}
in which one recognises  the \e{ weak value of the operator $\B$ for a system pre- and post-selected
in the states $|I\ra$ and $|f_j\ra$}, respectively (see, e.g., \cite{WVrev}). 
We will discuss the meaning of the weak values after studying the problem in a slightly different context.  
\section{Quasi-probabilities for virtual paths}
The classical analogy can be taken one step further. 
Quantum mechanics has no recipe for ascribing probabilities to interfering (virtual) alternatives, 
but one may try to obtain them by the  method which worked in the classical case of Sect.IV.
For  $\B=|b_i\ra\la b_i|\equiv \hat \pi _i$, one can measure the shift $Z_j(\hat \pi _i)$.
as well as  the net probability of the system reaching $|f_j\ra$ at $t_2$, $P^{S}(f_j\gets I)$. 
Their product should give something akin to the classical path probability in  Eq.(\ref{5b}). 
Indeed path  \e{probabilities}, defined in this manner,
   \begin{eqnarray} \label{1e}
\PP(f_j\gets b_i\gets I) \equiv 
\q\q\q\q\q\q\q\q\n 
\R\left [ \A^{S}(f_j\g b_i \g I)\sum_{i=1}^N\A^{S*}(f_j \g b_i\g I)\right ],\q\q
\end{eqnarray}
share  various properties 
with  their classical counterparts. In particular, they add up to one, 
   \begin{eqnarray} \label{2e}
\sum_{i,j=1}^N \PP(f_j\gets b_i\gets I)=\sum_{j=1}^NP^S(f_j\gets I)\q\n
 = \sum_{i=1}^NP^S(b_i\gets I)= 1,
\end{eqnarray}
their absolutes do not exceed unity,
$|\PP(f_j\gets b_i\gets I)|\le 1$, 
and the term multiplying $\delta(x_2-F_j)$ in Eq.(\ref{3d}) can be written as in the classical case [cf. Eqs.(\ref{2b}) and (\ref{3b}),  $\dx_1 \to \infty$]
 \begin{eqnarray} \label{3e}
P^{S}(f_j\gets I) \rho^I_1(x_1-\Z_j(\B))=\n
\sum_{i=1}^N \PP(f_j\gets b_i\gets I)\rho^I_1(x_1-B_i).
\end{eqnarray}
\newline
However, unlike their classical counterparts, {quasi-probabilities} in (\ref{1e})
can take negative values. 
Distributions which yield correct quantum mechanical averages, but may change sign,
are often used in quantum theory, the best known example being the Wigner functions \cite{Wig}.
It was Feynman who suggested that quantum mechanics can be formulated 
as a classical-like statistical theory  where probabilities are not restricted to be non-negative \cite{FeynN}. 
Equations (\ref{1e})-(\ref{2e}) clearly provide a simple example of this approach.
Next we consider a generic example. 
\section{The double-slit example}
Consider the \e{double-slit} case of $N=2$, where the states $|b_{1,2}\ra$
and 
$|f_{1,2}\ra$ play the roles of the \e{slits} and  \e{points on the screen}, respectively. 
The initial state can be chosen directed along the $z$-axis on the Bloch sphere, 
$|I\ra = |z_\up\ra$.  The other four are the states polarised up and down the directions $n$ and $n'$, whose 
azimuthal and polar angles are $(\phi,\theta)$ and $(\phi',\theta')$, respectively. 
For simplicity we assume that the system has no own dynamics, and put $\h^S=0$. 
The four paths are equipped with quasi-probabilities (\ref{1e}),
\begin{eqnarray}\label{f1}
\PP(n_\up'\gets n_\up\gets z_\up)= \cos^2 (\theta'/2)\cos^2(\theta/2)\q\q \n
+\cos(\phi-\phi')\sin (\theta)\sin (\theta')/4\equiv \PPP_1,\q\n
\PP(n'_\up\gets n_\dn \gets z_\up)= \cos^2 (\theta'/2)\sin^2(\theta/2)\q\q \n
-\cos(\phi-\phi')\sin (\theta)\sin (\theta')/4,\equiv \PPP_2\q\n
\PP(n_\dn'\gets n_\up\gets z_\up)= \sin^2 (\theta'/2)\cos^2(\theta/2)\q\q \n
-\cos(\phi-\phi')\sin (\theta)\sin (\theta')/4,\equiv \PPP_3\q\n
\PP(n'_\dn \gets n_\dn \gets z_\up)= \sin^2 (\theta'/2)\sin^2(\theta/2)\q\q \n
+\cos(\phi-\phi')\sin (\theta)\sin (\theta')/4\equiv \PPP_4,\q
\end{eqnarray}
at least one of which is likely to be negative. 
\newline
The regions where that happens are mapped in Fig.4 for $\phi'=0$, $\theta'=0.95\pi$
and various choices of the intermediate states $|n_\up\ra$ and $|n_\dn\ra$. 
In particular, for 
\begin{eqnarray}\label{f1a}
\phi=\pi, \q \theta=\pi/2\q
\end{eqnarray}
(the red dot on Fig.4) one finds
\begin{eqnarray}\label{f2}
\PPP_1=-0.0360 \q \text{(from $z_\up$ to $n'_\up$ via $n_\up$)}, \n
\PPP_2=0.0422 \q \text{(from $z_\up$ to $n'_\up$ via $n_\dn$)},\q\n
\PPP_3=0.5630 \q \text{(from $z_\up$ to $n'_\dn$ via $n_\up$)},\q\n
\PPP_4=0.4578\q \text{(from $z_\up$ to $n'_\dn$ via $n_\dn$)}.\q
\end{eqnarray}
where
$\PPP_1+\PPP_2= \cos^2(0.475 \pi)=P(n'_\up \g z_\up)$  and
$\PPP_3+\PPP_4= \sin^2(0.475 \pi)=P(n'_\dn \g z_\up)$.
\begin{figure}[h]
\includegraphics[angle=0,width=7.5cm, height=5cm]{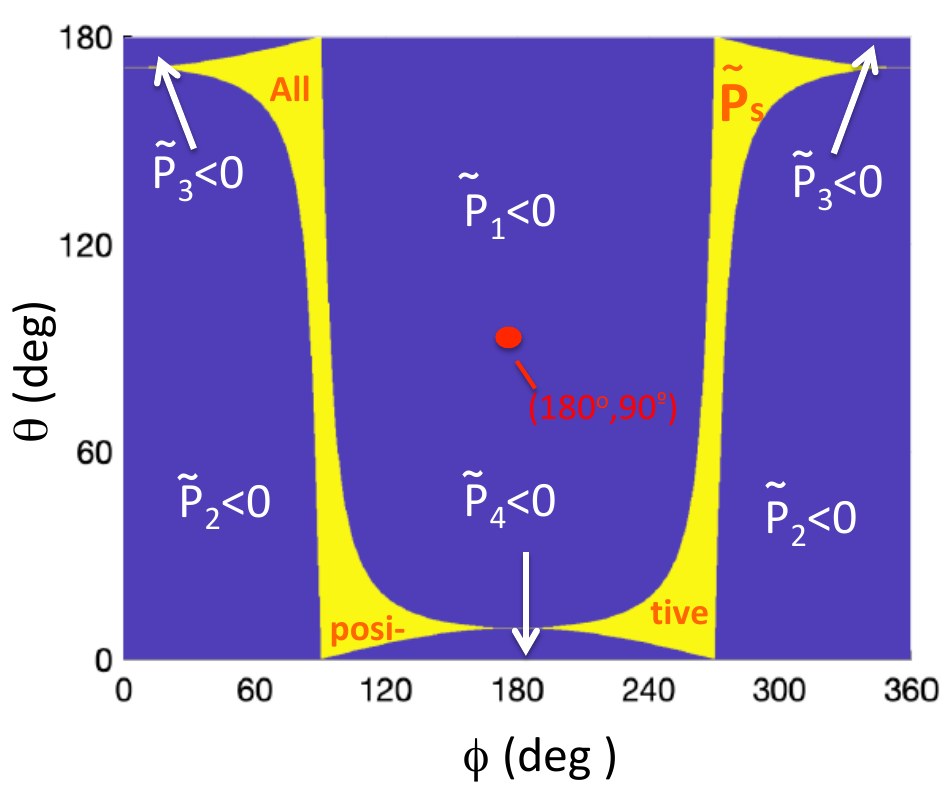}
\caption {The direction along which the intermediate states $|b_{1,2}\ra =|n_{\up,\dn}\ra$
are quantised is parametrised by its azimuthal and polar angles, $\phi$ and $\theta$.
The system $(N=2)$ is prepared in $|z_\up\ra$ and post-selected in $|n'_\up\ra $ or 
$|n'_\dn\ra $, where $\phi'=0$ and $\theta'=0.95\pi$. Inside the yellow region
all four path \e{probabilities} in Eqs.(\ref{f1}) are positive. Elsewhere, at least one of them 
has a negative value. }
\end{figure}
{\color{black}
\section{Interpretation}
\noindent
There appears to be a consensus about using negative conditional probabilities 
as intermediaries in calculating non-negative frequencies of observable physical events \cite{FeynN}, \cite{Bart}.
Such are the calculations of the observable probabilities $P^S(f_j\g I)$ and $P^S(b_i \g I)$ in Eqs.(\ref{2e})-(\ref{3e})
 (see also Appendix C). 
 \newline
 However \AA  can also measure $\hat \pi_i$, and observe the shift ${\PP(f_j\g b_i  \g I)}/{\sum_{i'}\PP(f_j\g b_{i'}  \g I) }<0$
 [cf. Eqs. (\ref{3e}) and (\ref{2})]. 
 This, however, should not be taken as an \e { experimental proof} that quantum probabilities, unlike their 
 classical counterparts, can be negative?}
 \newline
 The first objection is that a negative quantity cannot be  a probability, and must be something else.
 A probability is related to a frequency, i.e., the fraction of all cases,  in which the chosen scenario is realised.
 An observed number of cases cannot be negative, which is the main reason why a quantum system cannot be simulated 
 by a probabilistic classical computer \cite{FeynCOMP}.(see also Appendix D)
  \newline
Secondly, quasi-probabilities (\ref{1e})  are {\it non-local} quantities in the following sense.
Quasi-probability $\PP(f_j\gets b_i\gets I)$ 
depends [through the term $\sum_i A^{S*}(f_j \g b_i \g  I)$] on the amplitudes
of all paths, connecting $|I\ra$ with $|f_j\ra$. 
Thus, one hoping to probe the state of affairs in a particular path 
$f_j\gets b_i\gets I)$ by using an inaccurate
(or, which is the same, weakly coupled) pointer, discovers that his outcome depends 
on \e{what happens} in all paths interfering with the one he is interested in (see also Appendix E). 
{\color{black}
\section{Weak values and reshaping}
Thus, Alice has a set of \e{probabilities} which can change sign, 
each referring to the entire (sub)set of the paths leading to the same final outcome. 
As in the classical case [cf. Eq.(\ref{3b})] the observed shift of a highly inaccurate pointer 
can now be written as an \e{average},
\begin{eqnarray} \label{j1}
\Z_j(B) = \frac{\sum_{i=1}^N B_i\PP(f_j\g b_i\g I)}{\sum_{i=1}^N \PP(f_j\g b_i\g I)}.\q\q\q\q
\end{eqnarray}
For an unlikely transition, $P^S(j\g I) = \sum_{i} \PP(f_j\g b_i\g I) <<1$, 
the shift (\ref{j1}) can be made much larger than all the $B_i$s. 
It can be shown (see. e.g., \cite{DSS}) that one can always find $|I\ra$ and $|f_j\ra$
to give $\Z_j(B)$ any desired value between $-\infty$ and $\infty$. 
This, however, should not be taken as an 
\e {experimental proof}
that in the quantum world the value of any quantity can be, say, arbitrary large.} 
\newline
One only  needs to look in detail at the manner in which  such values are obtained.
In the classical case integration of Eq.(\ref{3b}) over $dx_2$ in the 
limit $\dx_1\to \infty$ yields  
[cf. Eqs.(\ref{4a}) and (\ref{3b})] 
\begin{eqnarray} \label{j2}
 \sum_{j=1}^N P(j\g I) \rho_1^I(x_1-z_j(B))=\rho_1^I(x_1-y(B)),\q\q 
\end{eqnarray}
where $z_j(B)$ is given by Eq.(\ref{3b}). 
It is easy to check that the causality principle (\ref{5b}) 
holds also in the quantum case, where one finds [cf. Eqs.(\ref{8c}) and (\ref{j1})]
\begin{eqnarray} \label{j3}
 \sum_{j=1}^N P^{S}(f_j\g I) \rho_1^I(x_1-\Z_j(B))=
\rho_1^I (x_1-Y(B)).\q\q
\end{eqnarray}
In both  cases the r.h.s. of Eqs. (\ref{j2}) and  (\ref{j3}) gives the distribution, obtained 
in a measurement without post-selection. In the l.h.s  this distribution 
is partitioned into sub-distributions, conditional on post-selection
in one of the final  states.
The only difference is that classically  the shift $z(B)$ always lies 
between $min (B_i)$ and $max(B_i)$, whereas in the quantum case $Z(B)$ are not restricted to this interval.
By post-selecting in a state $|f_j\ra$ \AA can separate the corresponding term in the l.h.s. 
of Eq,(\ref{j3}), a Gaussian with a large shift $\Z_j(B)$. The Gaussian is duly reduced 
by the factor $P^{S}(f_j\g I)$ which is small since 
the term 
must fit under the tail of a larger Gaussian $\rho_1^I (x_1-Y(B))$. 
\newline
One misunderstanding may arise if one assumes that  larger values 
of $x_1$ correspond also to larger values of $B$. 
This is, indeed, true in an accurate measurement, where 
the initial distribution $\rho^I\left (x_1\right)$ is narrow.
This is no longer true if $\dx_1 >> max(B_i)$,  
since a large reading $x_1>>\text{max}(B_i)$ may 
 come from any  of the broad Gaussians $\rho_1^I(x_1-B_i)$.
\newline
Indeed, in Eq.(\ref{j3})  each term in the l.h.s. of Eq.(\ref{j3}) 
can be obtained without post-selection, 
by first performing a two-point measurement of $B$, and then accepting a reading $x_1$ 
with a probability 
\begin{eqnarray} \label{j4}
\om_{j}(x_1) =\frac{P^{S}(j\g I) \rho^I(x_1-\Z_j(B))}
{\sum_{j=1}^N P^{S}(j\g I) \rho^I(x_1-\Z_j(B))}, 
\end{eqnarray}
so that
\begin{eqnarray} \label{j5}
\om_{j}(x_1)\rho_1^I\left (x_1-Y(B)\right)=P^{S}(j\g I) \rho_1^I(x_1-\Z_j(B)).\q\q
\end{eqnarray}
With all $P^S$ and $\rho_1^I$ non-negative, one finds $0\le \om_j(x_1) \le 1$ 
so such a {\it reshaping} is always possible.
Clearly, no anomalously 
large values of $B$ are selected or detected by either method. 
\newline
\newline
\section{The double slit  example (continued)}
As an example, consider a measurement of a projection of a spin-1/2, 
\begin{eqnarray} \label{h1}
\B=|n_\up\ra\la n_\up|- |n_\dn\ra\la n_\dn|, \q B_1=-B_2=1,
\end{eqnarray}
on the direction $n_\up$,  $\phi=\pi$, $\theta=\pi/2$,  post-selected  
along the direction $n'_\up$, $\phi'=0$, $\theta'=0.95 \pi$, 
discussed in Sect. VIII. 
\begin{figure}[h]
\includegraphics[angle=0,width=7.5cm, height= 5.5cm]{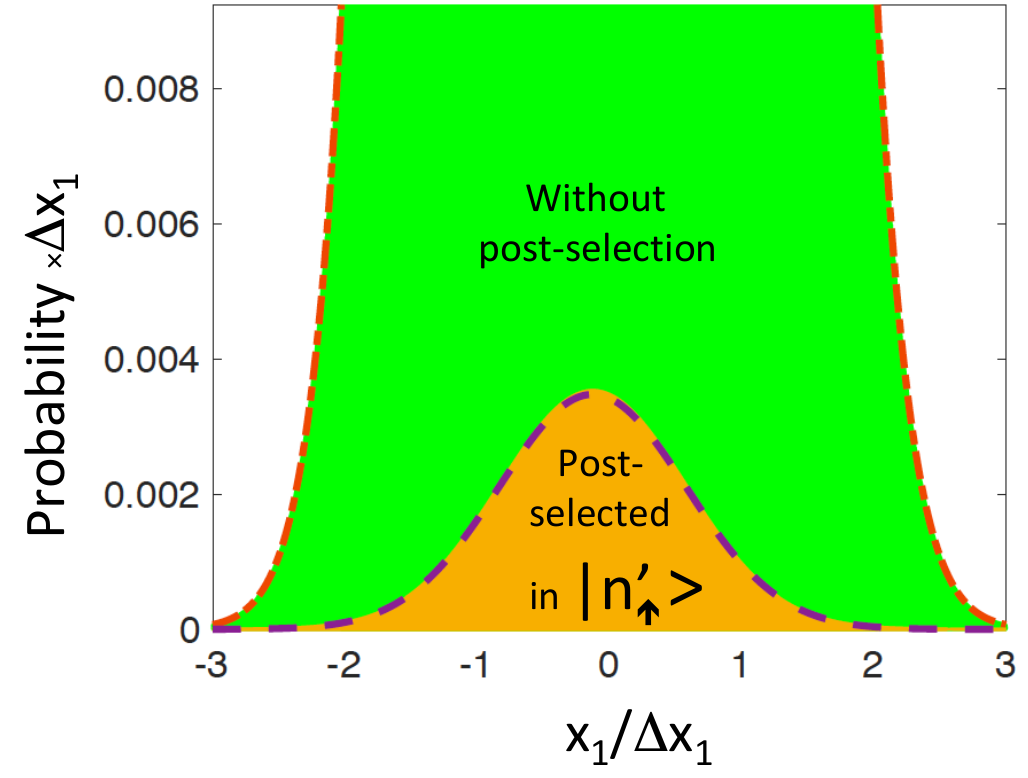}
\caption {a) Distributions of the readings of a highly inaccurate pointer, 
$\dx_1/(B_2-B1)=50$ measuring spin's projection onto the
direction $\phi=\pi$, $\theta=\pi/2$ .
The larger filled curve, centred at $x_1=0$,  is the distribution without post-selection (\ref{6c}), 
with the dot-dash line showing the approximation (\ref{4a}).
The smaller filled curve, centred at $x_1=-12.6051$ [cf. Eq.(\ref{h2})], is the unnormalised distribution [cf. Eq.(\ref{2d})] conditioned 
on post-selection in the state $|n'_\up\ra$ polarised along the 
direction $\phi'=0$, $\theta'=0.95\pi$.  The dashed line is the approximation 
(\ref{3d}). 
 }
\end{figure}
We have [cf Eqs. (\ref{1e}) and  (\ref{j1})]
\begin{eqnarray} \label{h2}
P^S(n'_\up \g z_\up)=\PPP_1+\PPP_2=0.0062, \n
Z_1=(\PPP_1-\PPP_2)/(\PPP_1+\PPP_2)=-12 .7062,\n
P^S(n_\up\g z_\up) =\PPP_1+\PPP_3 =0.5, \n
P^S(n_\dn\g z_\up) =\PPP_2+\PPP_4 =0.5, \n
B_1P^S(n_\up\g z_\up)+B_2 P^S(n_\dn\g z_\up)=0.
\end{eqnarray}
Results for $\dx_1/(B_1-B_2)=0.02$ are shown in Fig.4.
The distribution of the pointer's readings  without  post-selection
is a broad Gaussian centred at $x_1=0$.
The distribution, conditioned upon post-selection in $|n'_\up\ra$
is a much smaller Gaussian, centred at $x_1=-12.7062$. It fits, 
as it must,  under 
the larger curve, and can be obtained by \e{reshaping}, e.g.,  
by retaining only a fraction of the results,  with the acceptance probability 
(\ref{j4}) given by [$x_1\le 2009.5$ so $\om_1(x_1)\le 1$] \cite{FOOT},
\begin{eqnarray} \label{j4}
\om_{1}(x_1) \approx 0.0062 \exp(-0.0025 x-0.0161). 
\end{eqnarray}
\section{Conclusions}
In summary, in order to gain insight into  inaccurate (weak) quantum measurements 
we found it instructive to review a classical analogue first. 
Consider  a classical stochastic system which can reach a final  state $j$ by passing through an intermediate 
state $i$  where a variable $B$ takes a value $B_i$. The variable can be measured by moving a pointer 
 by $B_i$ units whenever the system passes through the $i$-th state. The pointer's initial position 
 is itself uncertain, and various distributions of its readings are obtained for each final state $j$.
Adding them one obtains the \e{unconditional} distribution for an experiment without \e{post-selection}.
This ensures that 
future observations may not alter the statistics already recorded. 
\newline
An accurate pointer yields information about the path taken in each trial. 
With a highly inaccurate pointer this \e{single-system} information is lost, 
yet one obtains such \e{ensemble}  information as the average of the measured $B$, 
or the individual path probabilities. 
\newline
The above analysis can be extended to the quantum case simply by allowing some
of the path {probabilities} to take negative values (see Sect. VII). 
As in the classical case, with post-selections made, the distributions of  a highly inaccurate pointer's readings
in Eq.(\ref{j3}) add up to the {unconditional} distribution (\ref{7c}). 
As in the classical case, all such distributions are (reduced) shifted copies of the initial one.
\newline
However, while classically all the shifts must lie between the smallest and the largest values of the measured $B$, 
there is no restriction on the values of  the quantum shifts in Eq.(\ref{j1}).
[This is why \e{anomalous weak values} erroneously predicted in \cite{Ferry} cannot exist in classical
statistics. They require path probabilities of opposite signs, and these do not exist in the classical case 
of Sects.III and IV.]
\newline
We note further that the shifts of a highly inaccurate (weak) quantum pointer 
can be seen as \e{ensemble property} of a bizarre statistical ensemble where the probabilities 
may take negative values. Such probabilities capture the essential faetures of 
quantum motion, such as an increase in the detection rate if one the of paths is blocked \cite{FeynL}. 
\newline
Although, it may look as if in the unlikely transition 
between the system's initial and final states a system's variable takes
a unique unexpectedly large value, this is not the case.
The large pointer's readings occur not because the pointer was 
pushed especially far when the post-selection was made. Rather they have been
selected from the readings already present there owing to the large size
of the initial distribution. Exactly the same effect can be achieved 
by simply retaining the larger readings of the distribution obtained without post-selection, and discarding the rest (see Sects. X and XI)
The notion of  uncommonly large, or otherwise \e{unusual}, physical values lurking in quantum transitions where
the interference between the relevant pathways is not destroyed is, therefore, spurious.
\newline
Finally, we note the similarity with the phenomenon of the apparent \e{superluminality} observed in quantum tunnelling (see e.g.,\cite{SL1}-\cite{SL2}). 
There the strongly reduced transmitted wave packet 
builds up from delayed copies of the freely propagating state \cite{DSsl}. Destructive interference ensures that the 
 wave packet lies ahead of the freely propagating one creating the impression of a particle 
crossing the barrier infinitely fast. However, one does not have to worry about violation of Einstein's 
relativity \cite{DScomm}, just as one does not have to worry about the unusually large spin of Sect. IX. 
Both indicate the expected failure of a measurement protocol designed to circumvent the Uncertainty Principle \cite{FeynL}.

\section*{Appendix A. Equations. (\ref{1}) and (\ref{3})}
A change of variables $y=x/\dx$ in Eq.(\ref{1}) yields
\begin{eqnarray} \label{A1}
\sum_{i=1} ^N A_iG(y-B_i/\dx) \approx  \sum_{i=1} ^N A_i[G(y)-\partial_yG(y)B_i/\dx]\q\q\n
\approx  \sum_{i=1} ^N A_I\left [G(y) -\partial_yG(y)\sum_{i=1} ^N A_iB_i/\dx \sum_{i=1} ^N A_i\right ]\n
\approx \sum_{i=1} ^N A_I\times G(y -z/\dx)= \sum_{i=1} ^N A_i\times G\left (\frac{x -z}{\dx}\right ).
\end{eqnarray}
Note that the third inequality holds if $|z|/\dx <<1$, which would require a large $\dx$ if $A_i$ 
have opposite signs, $\sum_{i} A_i$ is small, and $|z|$ is large. 
\newline
Furthermore, with complex $A_i$ one has $\left [ G(B_i,\dx)\equiv G\left(\frac{x-B_i}{\dx}\right )\right ]$
\begin{eqnarray} \label{6}
\left | \sum_{i=1}^N A_iG(B_i, \dx )\right |^2 = \left (\sum_{i=1}^N \R [ A_i]G(B_i, \dx )\right)^2+\q\q
\n \left (\sum_{i=1}^N \I [ A_i]G(B_i, \dx )\right)^2.
\end{eqnarray}
Applying  (\ref{3}) to each sum and adding up the results yields
\begin{eqnarray} \label{7}
\left | \sum_{i=1}^N A_iG(B_i, \dx )\right |^2 
{\xrightarrow[ \dx \to \infty ] {}}\left | \sum_{i=1}^N A_i\right |^2 \times \n
G^2\left (x,\frac{\R \left [\sum_{i=1}^N A_i^*\times 
\sum_{i=1}^N A_iB_i\right]  }{|\sum_{i=1}^NA_i|^2}, \dx \right ).
\end{eqnarray}
A simple calculation shows that the second argument of the $G$ in Eq.(\ref{6}) reduces to $\R \left [{\sum_{i=1}^N A_iB_i}/{\sum_{i=1}^NA_i}\right ]$. 
\newline
\section*{Appendix B. von Neumann pointers}
Alice can monitor a quantum system by means of pointers (P), 
heavy particles in one dimension,  whose positions she can observe directly.
The pointers are briefly coupled to the system, at $t_1$, and just before $t_2$. 
The Hamiltonian, therefore, is ($\hbar=1$)
\begin{eqnarray}\label{k1}
\h^{S+P_1+P_2}(t)= \h^S +\h^{S+P_1}(t) + \h^{S+P_2}(t).\q
\end{eqnarray}
The pointers are designed to measure operators $\B$ and $\F$, respectively, 
\begin{eqnarray}\label{k2}
 \h^{S+P_{1,2}}(t)=\hat p_{1,2}\times (\B,\F) \times \delta(t-t_{1,2}).
\end{eqnarray}
where $\hat p_l=\int dp_l |p_l\ra p_l \la p_l|$, $\la x_l|p_l\ra =\exp(ip_lx_l)$, 
$l=1,2$, is the $l$-th pointer's momentum operator. 
(There is also a third pointer, by means of which A. establishes
the system's initial state $|I\ra$, and which we omit).
For any  state $|\psi\ra$
one finds
 \begin{eqnarray}\label{k3}
 \exp\left[-i\int \h^{S+P_1}(t) dt\right ] |\psi\ra= 
 \n
 \sum_{i=1}^N \la b_i|\psi\ra |G_1(B_i)\ra |b_i\ra,\q\q\n
 |G_1(B_i)\ra\equiv \int dx_1\la x_1-B_i|G_1\ra |x_1\ra, \q
\end{eqnarray}
(and similarly for the second pointer).
Thus, initial state of the composite $|\Phi(0)\ra=|G_2\ra|G_1\ra|I\ra$ evolves into 
 \begin{eqnarray}\label{k4}
| \Phi(t_2)\ra= \sum_{i,j=1}^N |f_j\ra  |G_2(F_j)\ra| G_1(B_i)\ra\times
\n
 \la f_j| \u^S(t_2-t_1)
  |b_i\ra   \la b_i|   \u^S(t_1)  |I\ra,
\end{eqnarray}
and projecting it onto 
$|x_2\ra|x_1\ra |f_j\ra$ yields Eq.(\ref{3d}). 

\section*{Appendix C. Negative probability and observables}
In Alice's experiment with a highly inaccurate
 pointer, $\dx_1 \to \infty$, 
there are three directly observable probabilities.
One is defined by the rule (ii)  for the system in isolation [cf. Eq.(\ref{2e})],  
\begin{eqnarray}\label{g0}
P^S(f_j\g I)= \sum_i \PP_{ij}=|\la f_i|\u^s(t)|I\ra|^2 \ge 0,
\end{eqnarray}
where we write $\PP_{ij}=\PP(f_j\gets b_i\gets I)$.
The probability to find a reading $x_2$ in the vicinity of $F_j$ agrees with it
($\epsilon \to 0$), 
\begin{eqnarray}\label{g-1}
P(x_2=F_j)\equiv \int_{F_j-\epsilon}^{F_j-\epsilon} dx_2\int dx_1\rho(x_1,x_2)=\sum_i \PP_{ij}.\q\q
\end{eqnarray}
 Finally, there is the probability to find a reading $x_1$ given that $x_2$ is in the vicinity of $F_j$
 \begin{eqnarray}\label{g-1}
P(x_1,x_2F_j)\equiv \int_{F_j-\epsilon}^{F_j-\epsilon} dx_2\rho(x_1,x_2)=\n
\sum_i \rho^I_1(x_1-B_i)\PP_{ij}\ge 0.\q\q
\end{eqnarray}
Furthermore, negative probability of a scenario may mean that the scenario 
cannot be verified directly \cite{FeynN}. Indeed, in neither of the three cases the system 
is seen to travel one path and not the other. 
\newline
Finally, the same is true even when quasi-probabilities  of all paths
are non-negative (e.g,, in the light-coloured area in Fig.3),
One still cannot have the probabilities $\PP(f_j\gets b_i\gets I)=\R\left [ \A^{S}(f_j\g b_i \g I)\A^{S*}(f_j \g I)\right ]$, 
and know the path travelled by the system.
An attempt to distinguish between the paths by using an accurate pointer
would create different probabilities $P^S(f_j\gets b_i\gets I)=|\A^S(f_j\gets b_i\gets I)|^2$. 
\section*{Appendix D. An example}
Here is a simple illustration, 
 With the classical path probabilities $P(j\g i \g I)$ known, A. can simulate, e.g., the $P(i=3\g I)$ in Eq.(\ref{1a}).
 [E.g., by  splitting interval $[0,1]$ into $N^2$ segments, with each length proportional to $P(j\g i \g I)$.
Whenever random number generator yields a number within a segment, the ball travels the corresponding path.]
Dividing the number of cases in which the ball was seen, e.g.,  in the node $3$, $N_3$, by the total number of trials $N>>1$, 
Alice estimates $P(3\g I)\approx N_3/N$.] In the quantum case, where at least one  $\PP(F_j\g b_i  \g I)<0$,
the method fails,
since instruction to sample with a negative probability is obviously meaningless. 
\section*{Appendix E. Non-locality of quasi-probabilities}
In the simplest classical case $N=2$ Alice can alter the probability 
$P(1\g 2\g I)$, while keeping the $P(1\g 1\g I)$ intact simply by altering 
the branching probability at the second node, $P(1\g 2)\to P'(1\g 2)$. 
Now using the pointer which monitors the path $1\g 1\g I$, $B_i=\delta_{i1}$, 
and multiplying its shift by $P(1\g 1\g I)+P'(1\g 1\g I)$ she will always
recover the same path probability $P(1\g 2\g I)$. We call such measurement 
\e{local}, since A. result only depends on \e{what happens} in the path $1\g 1\g I$. 
\newline 
This is no longer true  in the quantum case, where a path is described by an 
amplitude, rather than a path probability. In the double-slit case of Sect.VIII
$N=2$,
with $\phi=\phi'$, the amplitudes for the paths leading to $|n'_\up\ra$ 
are given by 
\begin{eqnarray}\label{AC1}
\A(n_\up'\gets n_\up\gets z_\up)=\cos (\theta/2)\cos(\theta/2-\theta'/2),\q\q\n 
\A(n_\up'\gets n_\dn\gets z_\up)=\sin (\theta/2)\cos(\theta/2-\theta'/2),\q\q\n
\A(n_\up'\gets n_\up\gets z_\up)+\A(n_\up'\gets n_\dn\gets z_\up)=\cos(\theta'/2).
\end{eqnarray} 
For $\beta <1$ and a fixed $0<\theta< 2 \arccos(\beta/2)$, the  choice
of the final state
 \begin{eqnarray}\label{AC2}
\theta'= \theta + 2\arccos (\beta/\cos(\theta/2),
\end{eqnarray}
keeps $\A(n_\up'\gets n_\up\gets z_\up)=\beta$ constant, while
the second amplitude on Eq.(\ref{AC1}) and 
 \begin{eqnarray}\label{AC2}
\PP(n_\up'\gets n_\up\gets z_\up)=\beta\cos(\theta'/2),
\end{eqnarray}
both change. By using a highly inaccurate  pointer, which moves only when 
the system follows the path $n_\up'\gets n_\up\gets z_\up$, 
A. recovers a quantity which depends on the amplitudes of all 
paths leading to the same outcome. In an optical realisation of 
the {double slit} experiment the paths in different arms of a 
Mach-Zehnder interferometer can be well separated in space. 
Alice would not be wrong to describe this behaviour as \e{non-local}.
\section*{Appendix F. Causality in the quantum case}
Distribution of the first pointer's readings,
regardless of what reads the second pointer,
 is [cf. Eqs.(\ref{1d}) and (\ref{2d})]
 \begin{eqnarray}\label{AE1}
\rho_1(x_1)=\sum_{j=1} ^N\int dx_2 
|G_2(x_2-F_j)|^2\times\q\q \q\q \n
\left | \sum_{i=1}^N \A^{S}(f_j\g b_i\g I) 
G_1(x_1-B_i)\right |^2=
\q\q\q\n
\sum_{i'i=1}^N \left [ \sum_{j=1} ^N \A^{S*}(f_j\g b_{i'})\A^S(f_j\g b_{i})\right ]A^{S*}(b_{i'}\g I) \n
\times\A^S(b_{i}\g I)G^*(x_1-B_{i'})G(x_1-B_i).\q\q
\end{eqnarray}
Since $\sum_{j=1} ^N \A^{S*}(f_j\g b_{i'})\A^S(f_j\g b_{i})=\la b_{i'}|b_i\ra =\delta_{ii'}$,
this reduces to 
 \begin{eqnarray}\label{AE2}
\sum_{i=1}^N|\A^S(b_{i}\g I)|^2|G(x_1-B_i)|^2=\q\q\n
 \sum_{i=1}^NP^S(b_i\g I)\rho_1^I(x_1-B_i),\q\q\q\q
 \end{eqnarray}
 which is the distribution (\ref{6c}) one obtains if no post-selection is made.
\end{document}